\documentstyle[sprocl,epsfig]{article}

\def\EPJ{{\em Eur. Phys. J.} C }

\def\be{\begin{equation}}
\def\ee{\end{equation}}
\def\bea{\begin{eqnarray}}
\def\eea{\end{eqnarray}}

\begin{document}

\title{STUDYING HIGGS BOSONS BY TOP PAIR PRODUCTION AT PHOTON COLLIDERS
\footnote{based on the work in collaboration with 
\uppercase{K}.~\uppercase{H}agiwara~[1].}\\}

\author{ ERI ASAKAWA }

\address{Theory Group, KEK, Tsukuba, Ibaraki 305-0801, Japan}


\maketitle\abstracts{
We study the effect of heavy neutral Higgs bosons on the 
$t\overline{t}$ production process at photon linear colliders.
The interference patterns between the resonant Higgs production
amplitudes and the continuum QED amplitudes are examined.
The patterns are sensitive to the phase of the $\gamma\gamma$-Higgs vertex
which is a good probe of new charged particles.
}

\section{Introduction}
New physics contribute to the Higgs production
at photon colliders via radiative corrections from the new charged particles. 
When the photon-photon-Higgs vertex is written by
\begin{eqnarray}
{\it L}_{\phi \gamma \gamma} &=& \frac{1}{m_\phi}
\left( b_\gamma^H A_{\mu\nu} A^{\mu\nu} +
b_{\gamma}^A \widetilde{A}_{\mu\nu} A^{\mu\nu} \right) \phi,
\end{eqnarray}
where $\phi$ denotes a spinless boson
$H$ or $A$ ($H$ and 
$A$ are the CP-even and CP-odd Higgs bosons respectively.),
the $\gamma\gamma\phi$ vertex parameter $b_\gamma^\phi$
generally has complex phase even in CP invariant theories of
the Higgs sector.
%
%
Because the imaginary part
of $b^\phi_\gamma$ is a sum of
the contribution from the $\phi$ decay modes into charged particles
whereas the real part receives contribution from all the charged 
particles, we expect that $\arg(b_\gamma^\phi)$ is a good probe
of heavy charged particles.

In this report, we study the interference patterns of the resonant 
and the continuum amplitudes
for the $\gamma\gamma \rightarrow
t \overline{t}$ process by using the circularly 
polarized colliding photons~[1]. 
It will be shown that
these interference effects allow us to observe the complex phase of 
the $\gamma\gamma$-Higgs vertices, as has been shown
in $WW$ and $ZZ$ production processes~[2].

\section{Observables For The Process $\gamma \gamma 
\rightarrow t \overline{t}$}

The helicity amplitudes for the process $\gamma_{\lambda_1}
\gamma_{\lambda_2} \rightarrow
t_\sigma \overline{t}_{\overline{\sigma}}$ can be expressed as
\begin{eqnarray}
\label{ampeq}
{\it M}_{\lambda_1 \lambda_2}^{\sigma \overline{\sigma}} =
\left[ {\it M}_\phi \right]
_{\lambda_1 \lambda_2}^{\sigma \overline{\sigma}} +
\left[ {\it M}_t \right]
_{\lambda_1 \lambda_2}^{\sigma \overline{\sigma}} ,
\end{eqnarray}
where the first term ${\it M}_\phi$ stands for the $s$-channel
$\phi$-exchange amplitudes and the latter term ${\it M}_t$
stands for the $t$- and $u$-channel top-quark-exchange amplitudes.
By considering the decay angular distribution of $t\overline{t}$
pairs, we can derive 
the convoluted
four observables, $\Sigma_1$ to $\Sigma_4$,
\begin{eqnarray}
&\Sigma_i&(\sqrt{s}_{\gamma\gamma})
\\ \nonumber
&=&
\int d\sqrt{s}_{\gamma\gamma} \sum_{\lambda_1,~\lambda_2}
\left( \frac{1}{{\it L}_{0.8}} 
\frac{d{\it L}^{\lambda_1\lambda_2}}
{d\sqrt{s}_{\gamma\gamma}} \right) \left( \frac{3\beta}
{32\pi s_{\gamma\gamma}}
\int S^i_{\lambda_1 \lambda_2} (\Theta, \sqrt{s}_{\gamma\gamma})d\cos\Theta
\right),
\end{eqnarray}
for $i=1-4$ where the functions $S_{\lambda_1 \lambda_2}^i$ contain
all the information about the $\gamma\gamma \rightarrow 
t\overline{t}$ helicity amplitudes:
\begin{eqnarray}
\label{S1toS4}
&&S^1_{\lambda_1 \lambda_2} = 
\left| {\it M}_{\lambda_1\lambda_2}^{RR} \right|^2,~~~~~~
S^2_{\lambda_1 \lambda_2} = \left|{\it M}_{\lambda_1\lambda_2}^{LL}\right|^2,
\\ \nonumber
&&S^3_{\lambda_1 \lambda_2} =
2\Re\left[{\it M}_{\lambda_1\lambda_2}^{RR} \left({\it M}
_{\lambda_1\lambda_2}^{LL}\right)^*\right],~~~
S^4_{\lambda_1 \lambda_2} =
2\Im\left[{\it M}_{\lambda_1\lambda_2}^{RR} \left({\it M}
_{\lambda_1\lambda_2}^{LL}\right)^*\right].
\end{eqnarray}
$\Theta$ is the polar angle of the top momentum in the
$\gamma\gamma$ CM frame, and
the normalized luminosity distribution for each
$\gamma\gamma$ helicity combination is expressed by 
$(1/{\it L}_{0.8})
d{\it L}^{\lambda_1 \lambda_2}/d \sqrt{s}_{\gamma\gamma}$
where ${\it L}_{0.8} \approx 0.1 {\it L}_{ee}^{geom}$.
In this report, the luminosity distribution is derived
by assuming $\sqrt{s}_{ee}=500$ GeV, $x=4.8$,
$P_l=-1.0$ and $P_e=0.9$.

\section{Effects Of The 
$\gamma\gamma\phi$ Phase On The Observables}

\begin{figure}[ht] 
     \begin{center}
     \vspace*{.2cm}
     \epsfig{file=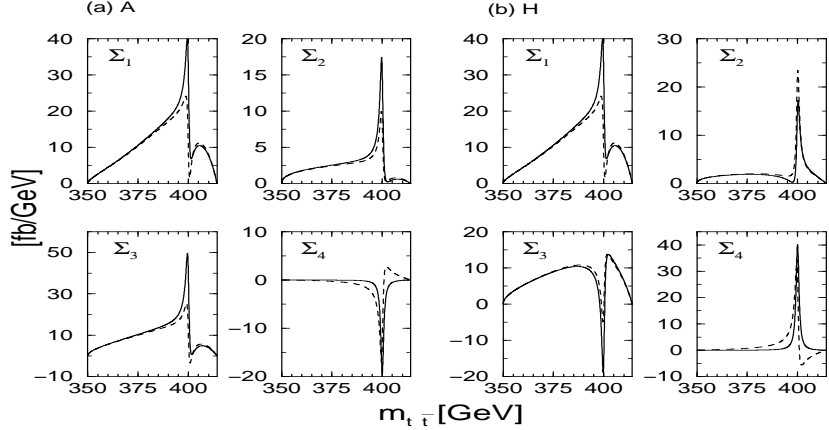,width=11.1cm,height=5.7cm}
     \end{center}
     \caption{The observables $\Sigma_1$ to $\Sigma_4$.
The solid and dashed curves indicate the case of
$\arg(b_\gamma^\phi)=0$  and $\pi/4$, respectively.
The observables with the $A$ boson production are in the left
figures whereas those with the
$H$ boson production are shown in the right.}\label{phase} 
     \end{figure}

We study the $\arg(b_\gamma^\phi)$
dependence of the four observables defined in the previous
section. These observables are sensitive to the interference
effects between the resonant $\phi$-production and
the continuum QED amplitudes.
Fig.~{\ref{phase}} shows the observables
$\Sigma_1$ to $\Sigma_4$ for the $A$ boson production in the left,
and for the $H$ boson production in the right-hand side.
We adopt a MSSM prediction for calculating
the magnitudes of the $\phi$-production amplitudes.
The MSSM parameters used here are as follows:
$m_A=400$ GeV, $\tan\beta=3$,
$m_{\widetilde{f}}=1$ TeV, $M_2=500$ GeV, $\mu=-500$ GeV.
The solid and dashed curves indicate the case of
$\arg(b_\gamma^\phi)=0$ and $\pi/4$, respectively.

When we compare the $\arg(b_\gamma^A)=0$ observables
and the $\arg(b_\gamma^A)=\pi/4$ observables
in Fig.~{\ref{phase}}(a), 
we notice that the magnitudes of all the
observables decrease, that is, the negative interference
becomes stronger for $\arg(b_\gamma^A)=\pi/4$.
This statement can also be applied to the observables for
the $H$ boson production in Fig.~{\ref{phase}}(b) except for $\Sigma_2$.
As to $\Sigma_2$, the magnitude increases and
the positive interference effect is enhanced for $\arg(b_\gamma^H)=\pi/4$,
because of the sign relation for the $\phi$-production amplitudes
which is useful for probing CP parity of Higgs bosons~[3];
$\left[ {\it M}_H \right]_{\lambda \lambda}^{LL}
= - \left[ {\it M}_H \right]_{\lambda \lambda}^{RR}$
while
$\left[ {\it M}_A \right]_{\lambda \lambda}^{LL}
= \left[ {\it M}_A \right]_{\lambda \lambda}^{RR}$.

\section{Conclusion}
We have studied the effects of heavy Higgs bosons in $t\overline{t}$
production process at photon colliders. We have introduced observables
by considering
the angular correlation of decay products of top quarks, and
found that the $\arg(b_\gamma^\phi)$ dependence of
the four observables are significant enough that the phase
of the $\gamma\gamma\phi$ vertex function may be measured
experimentally by a careful study of all the observables.


\section*{References}
\begin{thebibliography}{99}
\bibitem{ah}{E. Asakawa and K. Hagiwara \EPJ \rm {\bf 31} (2003) 351.}\\
\bibitem{nzk}{P. Niezurawski, A.F. Zarnecki and M. Krawczyk 
{\em JHEP} \rm {\bf 0211} (2002) 034.}\\
\bibitem{aksw}{E. Asakawa, J. Kamoshita, A. Sugamoto and I. Watanabe 
\EPJ \rm {\bf 14} (2000) 335.}\\
\end{thebibliography}
\end{document}